\documentclass[final]{aipproc}
\layoutstyle{8x11single}
\usepackage{epsf,graphics,amssymb,amsfonts,amsmath,array}

\newcommand{\bfnabla}{\mbox{\boldmath $\nabla$}}

\newcommand{\lQ}{\Lambda_{\rm QCD}}

\newcommand{\als}{\alpha_{\rm s}}

\newcommand{\siml}{{\ \lower-1.2pt\vbox{\hbox{\rlap{$<$}\lower6pt\vbox{\hbox{$\sim$}}}}\ }} 
\newcommand{\simg}{{\ \lower-1.2pt\vbox{\hbox{\rlap{$>$}\lower6pt\vbox{\hbox{$\sim$}}}}\ }}
\newcommand{\nn}{\nonumber}

\newcommand{\be}{\begin{equation}} 
\newcommand{\ee}{\end{equation}}
\newcommand{\bea}{\begin{eqnarray}} 
\newcommand{\eea}{\end{eqnarray}}
\newcommand{\beq}{\begin{equation}}
\newcommand{\eeq}{\end{equation}}
\newcommand{\bqa}{\begin{eqnarray}}
\newcommand{\eqa}{\end{eqnarray}}

\def \cf {C_F}

\def \nc {N_c}

\def \bfnabla {\boldsymbol{\nabla}}

\def \als {\alpha_{\mathrm{s}}}

\def \m2   {\mu^{2 \epsilon}}
\def \br {\mathbf{r}}

\begin{document}

\title{Quarkonium dissociation in a thermal bath}


\author{Antonio Vairo}{address={Physik-Department, Technische Universit\"at M\"unchen, James-Franck-Str. 1, 85748 Garching, Germany}}

\begin{abstract}
In an effective field theory framework we review the two main mechanisms of quarkonium dissociation in a weakly coupled thermal bath.
\end{abstract}

\maketitle

\section{Introduction}
Heavy particles may be sensitive to \emph{new fundamental degrees of freedom}
and serve as probes of new physics. Heavy particles themselves may appear as 
new fundamental degrees of freedom, an example discussed at this conference being
heavy Majorana neutrinos~\cite{Biondinihere}.
Heavy particles, however, may also serve as probes of \emph{new phenomena} emerging 
from particularly complex environments. We will focus on this second aspect here and 
discuss, in some special cases, how heavy quarkonia are affected by the hot medium 
formed in present-day heavy-ion colliders. 
At this conference, this has also been discussed elsewhere, for instance, in~\cite{Escobedohere}.

We consider a particle of mass $M$ heavy, if $M$ is much larger than 
any other energy scale $E$ in the system. In particular, if $M$ is also much larger than the momentum 
of the particle, this qualifies the particle as non-relativistic. 
The hierarchy of energy scales, $M \gg E$, allows an \emph{effective field theory} (EFT) treatment at a scale $\mu$ such that $M \gg \mu \gg E$. 
The EFT describes the dynamics of the effective degrees of freedom that exist at the scale $\mu$. 
These are the field $H$ representing the low-energy fluctuations of the heavy particle, and all other low-energy fields.
In a reference frame where the heavy particle is at rest up to fluctuations of order $E$ or smaller, the EFT Lagrangian has the form 
\be
{\cal L}  = H^\dagger i D_0 H + [\hbox{operators of dimension \,} d>4] \, \times\, \frac{1}{M^{d-4}}  + {\cal L}_{\rm light},
\label{genlag}
\ee
where ${\cal L}_{\rm light}$ describes all low energy fields besides $H$.
In the heavy-particle sector, the Lagrangian is organized as an expansion in $1/M$.
Contributions of higher-order operators to physical observables are suppressed by powers of $E/M$.
The EFT Lagrangian may be computed setting $E=0$.

A special case is the case of a heavy particle of mass $M$ in a medium characterized by a temperature $T$ such that  
$M \gg T$. The system is described at a scale $\mu$ such that $M \gg \mu \gg T$ by an EFT Lagrangian that has 
the same structure as \eqref{genlag} if the heavy particle is at rest up to fluctuations of order $T$ or smaller.
This follows from the fact that the Lagrangian may be computed setting $T=0$.  
In particular, the Wilson coefficients encoding the high-energy modes may be computed in vacuum.
The Lagrangian is again organized as an expansion in $1/M$, whose higher-order 
contributions to physical observables are suppressed by powers of $T/M$.

Temperature is introduced via the partition function. 
Sometimes it is useful to work in the \emph{real-time formalism}.
Despite the fact that in real time the degrees of freedom double (``1'' and ``2''), 
the advantages are that the framework becomes very close to the one of $T=0$ EFTs 
and that, in the heavy-particle sector, the second degrees of freedom, labeled ``2'', 
decouple from the physical degrees of freedom, labeled ``1''.
This often leads to a simpler treatment with respect to alternative calculations 
in imaginary time formalism~\cite{Brambilla:2008cx}.

In the following, we will consider the special case of a heavy quarkonium, i.e., a bound state 
of a heavy quark $Q$ and a heavy antiquark $\bar{Q}$, weakly bound, interacting with a weakly coupled plasma.
This means that we consider a quarkonium sufficiently tight to be described as a \emph{Coulombic bound state}
(e.g. the bottomonium ground state) interacting with a medium sufficiently hot to behave 
as a \emph{weakly coupled quark-gluon plasma}~\cite{Brambilla:2004wf,Brambilla:2010cs}.
We will compute thermal corrections to the width induced by the medium.  
We will call these, the quarkonium thermal width, $\Gamma$.

\section{Heavy quarkonia in a thermal bath}
Lattice data suggest a crossover from hadronic matter to a plasma of deconfined quarks and gluons 
happening at a \emph{critical temperature} $T_c = 154 \pm 9$ MeV~\cite{Bazavov:2014pvz}.
High energy densities with temperatures larger than $T_c$ are produced in heavy-ion collisions at RHIC and LHC.
High-energy probes are needed to identify and qualify the state of matter that is formed there.
Heavy quarkonium has been suggested as one of those probes~\cite{Matsui:1986dk}.
This is because heavy quarks are formed early in heavy-ion collisions, 
heavy quarkonium formation will be sensitive to the medium and the dilepton signal makes an ideally clean experimental probe.

\begin{figure}[htb]
\makebox[0truecm]{\phantom b}{\epsfxsize=11truecm \epsfbox{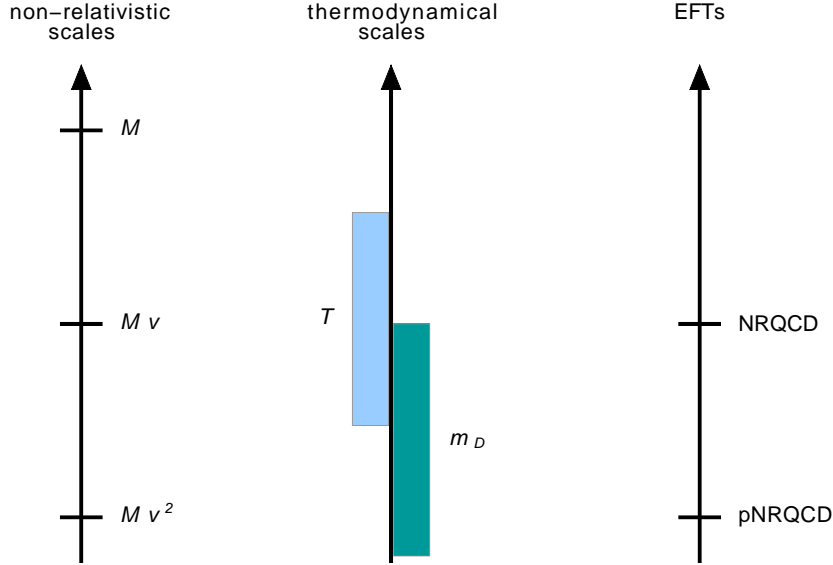}}
\caption{Energy scales and EFTs relevant for a heavy quarkonium in a weakly-coupled quark-gluon plasma.}
\label{figscales}
\end{figure}

Quarkonium, being a composite system, is characterized by several energy scales~\cite{Brambilla:2004jw}: 
$M$, $Mv$ (momentum transfer, typical inverse distance), 
$Mv^2$ (kinetic energy, binding energy $E_{\rm binding}$, potential $V$), ..., 
where $v$ is the relative heavy-quark velocity; $v\sim\als$ for a Coulombic bound state.
These in turn may be sensitive to thermodynamical scales smaller than the temperature.
The thermodynamical scales are: $\pi T$, $m_D$ (Debye mass, i.e. inverse screening length of the chromoelectric interactions), ... .
The non-relativistic scales are hierarchically ordered: $M \gg Mv \gg Mv^2$.
In a weakly-coupled plasma one assumes this also to be the case for the thermodynamical scales: $\pi T \gg m_D \sim gT$.
A weakly coupled quarkonium is the bottomonium ground state, $\Upsilon(1S)$;
produced in heavy-ion experiments at the LHC it may possibly realize the hierarchy of energy scales~\cite{Brambilla:2010vq,Vairo:2010bm}: 
$M_b \approx 5 \; \hbox{GeV} \; > M_b\als \approx 1.5 \; \hbox{GeV} \; >  \pi T \approx 1 \; \hbox{GeV} \; 
>  M_b\als^2 \approx 0.5  \; \hbox{GeV} \; \sim  m_D \simg \lQ$, for a temperature of the plasma of about $320$ MeV $\approx 2 T_c$.
The existence of a hierarchy of energy scales calls for a description 
of the system (quarkonium at rest in a thermal bath) in terms of a hierarchy of EFTs.
These are listed on the right arrow of Figure~\ref{figscales}.

\emph{Non-relativistic QCD} (NRQCD) is obtained by integrating out modes associated with the scale $M$~\cite{Caswell:1985ui,Bodwin:1994jh}
and possibly with thermodynamical scales larger than $Mv$. 
The Lagrangian is organized as an expansion in $1/M$ and is of the form~\eqref{genlag}:
\be
{\cal L}  =  \psi^\dagger \left( i D_0 + \frac{{\bf D}^2}{2 M} + \dots  \right) \psi   
+ \chi^\dagger \left( i D_0 - \frac{{\bf D}^2}{2 M} + \dots  \right) \chi  + \dots  + {\cal L}_{\rm light},
\label{NRQCDlag}
\ee
where $\psi$ ($\chi$) is the field that annihilates (creates) the (anti)fermion.

\emph{Potential non-relativistic QCD} (pNRQCD) is obtained by integrating out modes associated 
with the scale $Mv$~\cite{Pineda:1997bj,Brambilla:1999xf} and possibly with thermodynamical scales larger than $Mv^2$. 
The degrees of freedom of pNRQCD are quark-antiquark states (color singlet S, color octet O), 
low energy gluons and light quarks propagating in the medium.
The Lagrangian is organized as an expansion in $1/M$ and $r$, the distance between the heavy quark and antiquark, 
\bea
{\cal L} &=& \int d^3r \; {\rm Tr}\, \left\{ 
   {\rm S}^\dagger \left( i\partial_0 +  \frac{{\bfnabla_r}^2}{M} -  V_s + \dots \right){\rm S} 
+  {\rm O}^\dagger \left( i{D_0}      +  \frac{{\bfnabla_r}^2}{M} -  V_o + \dots \right){\rm O}\right\}
\nn\\
&&
\hspace{7.1mm}
+ {\rm Tr} \left\{  {\rm O^\dagger}{\bf r}\cdot g {\bf E}\,{\rm S} + {\rm H.c.} \right\} 
+ \frac{1}{2} {\rm Tr} \left\{  {\rm O^\dagger}{\bf r} \cdot g {\bf E}\,{\rm O}+ {\rm c.c.} \right\} + \cdots + {\cal L}_{\rm light},
\label{pNRQCDlag}
\eea
where $V_s$ and $V_o$ are the quark-antiquark colour-singlet and colour-octet potentials respectively, and ${\bf E}$ is the chromoelectric field.
The term ${\cal L}_{\rm light}$ in the NRQCD and pNRQCD Lagrangians describes the (in-medium) propagation of gluons and light quarks.

\subsection{Dissociation}
A quantity that may be relevant for describing the observed quarkonium dilepton signal 
emerging from heavy-ion collisions is the \emph{quarkonium thermal dissociation width}.
Two distinct dissociation mechanisms may be identified at leading order:
\emph{gluodissociation}, which is the dominant mechanism for temperatures such that $m_D \ll Mv^2$, 
\emph{dissociation by inelastic parton scattering}, which is the dominant mechanism for temperatures such that $m_D \gg Mv^2$.
Beyond leading order the two mechanisms become intertwined and a distinction between them less practical.

\begin{figure}[htb]
\makebox[-3.4truecm]{\phantom b}\put(0,0){\epsfxsize=7truecm \epsfbox{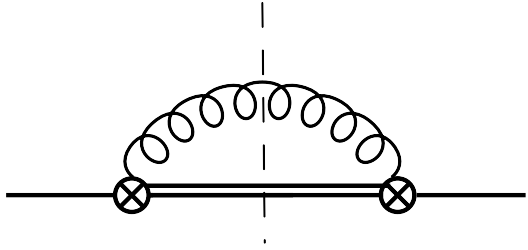}}
\caption{pNRQCD cutting diagram responsible for gluodissociation at leading order. 
In a pNRQCD Feynman diagram, single lines stand for quark-antiquark colour singlet propagators, 
double lines for colour octet propagators, curly lines for gluons and a circle with a cross for a chromoelectric dipole vertex.}
\label{figgluo}
\end{figure}

Gluodissociation is the dissociation of quarkonium induced by the absorption of a gluon from the medium~\cite{Kharzeev:1994pz,Xu:1995eb}.
The exchanged gluon is lightlike or timelike. The process happens when the gluon has an energy of order $Mv^2$,
therefore it may be described at the level of pNRQCD by the cutting diagram shown in Figure~\ref{figgluo}.
For a quarkonium at rest with respect to the medium, the dissociation width can be written as 
\be
\Gamma_{nl}= \int_{q_\mathrm{min}}\frac{d^3q}{(2\pi)^3}\,n_{\rm B}(q)\,\sigma_{\rm gluo}^{nl}(q),
\label{dissociationwidth}
\ee
where $\sigma_{\rm gluo}^{nl}$ is the in-vacuum cross section $(Q\overline{Q})_{nl} + g \to Q + \overline{Q}$, 
and $n_{\rm B}$ the Bose--Einstein distribution.
Gluodissociation is also known as \emph{singlet-to-octet break up}~\cite{Brambilla:2008cx}.

\begin{figure}[htb]
\makebox[-8truecm]{\phantom b}\put(0,0){\epsfxsize=16truecm \epsfbox{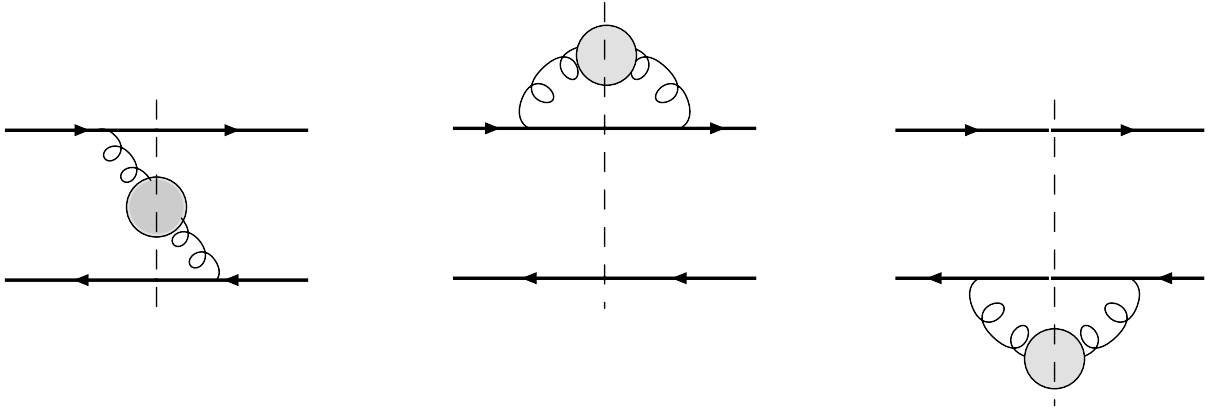}}
\caption{NRQCD cutting diagrams responsible for dissociation by inelastic parton scattering at leading order.
In a NRQCD Feynman diagram, single lines stand for quark or antiquark propagators.  
The shaded blobs represent gluon self-energy diagrams.}
\label{figpar1}
\end{figure}

\begin{figure}[htb]
\includegraphics[height=.15\textheight]{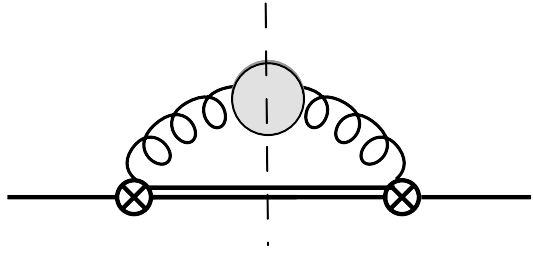}
\caption{pNRQCD cutting diagram responsible for dissociation by inelastic parton scattering at leading order.}
\label{figpar2}
\end{figure}

Dissociation by inelastic parton scattering is the dissociation of quarkonium by
scattering with gluons and light-quarks in the medium~\cite{Grandchamp:2001pf,Grandchamp:2002wp}. 
The exchanged gluon is spacelike. In the NRQCD Lagrangian each transverse gluon is suppressed by $T/M$ (see \eqref{NRQCDlag}), 
hence at leading order one just needs to consider diagrams with one exchanged gluon.
If the exchanged momentum is of order $Mv$, then dissociation by inelastic parton scattering
is described at the level of NRQCD by the cutting diagrams shown in Figure~\ref{figpar1}.
If the exchanged momentum is of order $Mv^2$, then it is described at the level of pNRQCD by the cutting diagram shown in Figure~\ref{figpar2}.
For a quarkonium at rest with respect to the medium, the thermal width has the form~\cite{Brambilla:2013dpa}
\be
\Gamma_{nl}=\sum_p\int_{q_\mathrm{min}}\frac{d^3q}{(2\pi)^3}\,f_p(q)\,\left[1\pm f_p(q)\right]\,\sigma_p^{nl}(q),
\label{partonwidth}
\ee
where the sum runs over the different incoming and outcoming light partons $p$ 
($p=g$ for gluons with the plus sign, and $p=q$ for quarks with the minus sign),  
$f_g(q)=n_{\rm B}(q)=1/(e^{q/T}-1)$ and $f_q(q)=n_{\rm F}(q)=1/(e^{q/T}+1)$. 
The cross section $\sigma_p^{nl}$ may be identified with the in-medium cross section $(Q\overline{Q})_{nl} + p \to  Q + \overline{Q} + p$.
We observe that the formula differs from the gluodissociation formula \eqref{dissociationwidth}, 
a point not always appreciated by the literature on the subject.
Dissociation by inelastic parton scattering is also known as \emph{Landau damping}~\cite{Laine:2006ns}.

We define the quarkonium \emph{dissociation temperature} as the temperature at which $\Gamma \sim E_{\rm binding}$ 
and the \emph{screening temperature} as the temperature at which $1/m_D$ is of the order of the 
quarkonium Bohr radius. At weak coupling, it holds that $\pi T_{\rm dissociation} \sim M g^{4/3}$ and 
$\pi T_{\rm screening} \sim M g$. This implies that $T_{\rm screening} \gg T_{\rm dissociation}$.
In the special case of the $\Upsilon(1S)$, $T_{\rm dissociation}$ has been estimated to be about 450 MeV~\cite{Escobedo:2010tu}, 
which implies that $T \approx$ 320 MeV is below the dissociation temperature. 

In the following, we will consider different temperature regimes below the dissociation (and hence the screening) temperature
and above the critical temperature~\cite{Brambilla:2013dpa}.
For these regimes we will provide the dissociation cross sections for a quarkonium that is a Coulombic $1S$ bound state.
The wave function is $\langle\br\vert1S\rangle=1/(\sqrt{\pi}a_0^{3/2})\exp(-r/a_0)$, $a_0=2/(M \cf \als)$ the Bohr radius, 
$\cf = (\nc^2-1)/(2\nc) = 4/3$ and $\nc=3$ the number of colours.

\subsubsection{The temperature region $T\gg Mv\gg m_D$}
In the temperature region $T\gg Mv\gg m_D$ the $1S$ dissociation cross section by parton scattering is 
\be
\sigma^{1S}_{p}=\sigma_{cp}\,f(m_Da_0),
\ee
where 
\bea 
&& \sigma_{cq}\equiv8\pi C_Fn_f\,\als^2\,a_0^2, 
\\
&& \sigma_{cg}\equiv8\pi C_F\nc\,\als^2\,a_0^2, 
\\
&& f(m_Da_0) \equiv \frac{2}{(m_Da_0)^2}\left[1-4\frac{(m_Da_0)^4 -16 + 8 (m_Da_0)^2\,\ln\left(4/(m_Da_0)^2\right)}{((m_Da_0)^2-4)^3}\right],
\eea
and $n_f$ is the number of light quarks.
The cross section is not related, even at leading order, with a zero temperature process. 
The underlying reason is the infrared (IR) sensitivity of the cross section at the momentum scale $Mv$. 
This IR sensitivity is cured by the hard-thermal loop resummation at the scale $m_D$ and signaled by the 
logarithm $\ln (m_Da_0)$. The cross section is momentum independent.

\begin{figure}[htb]
\includegraphics[height=.32\textheight]{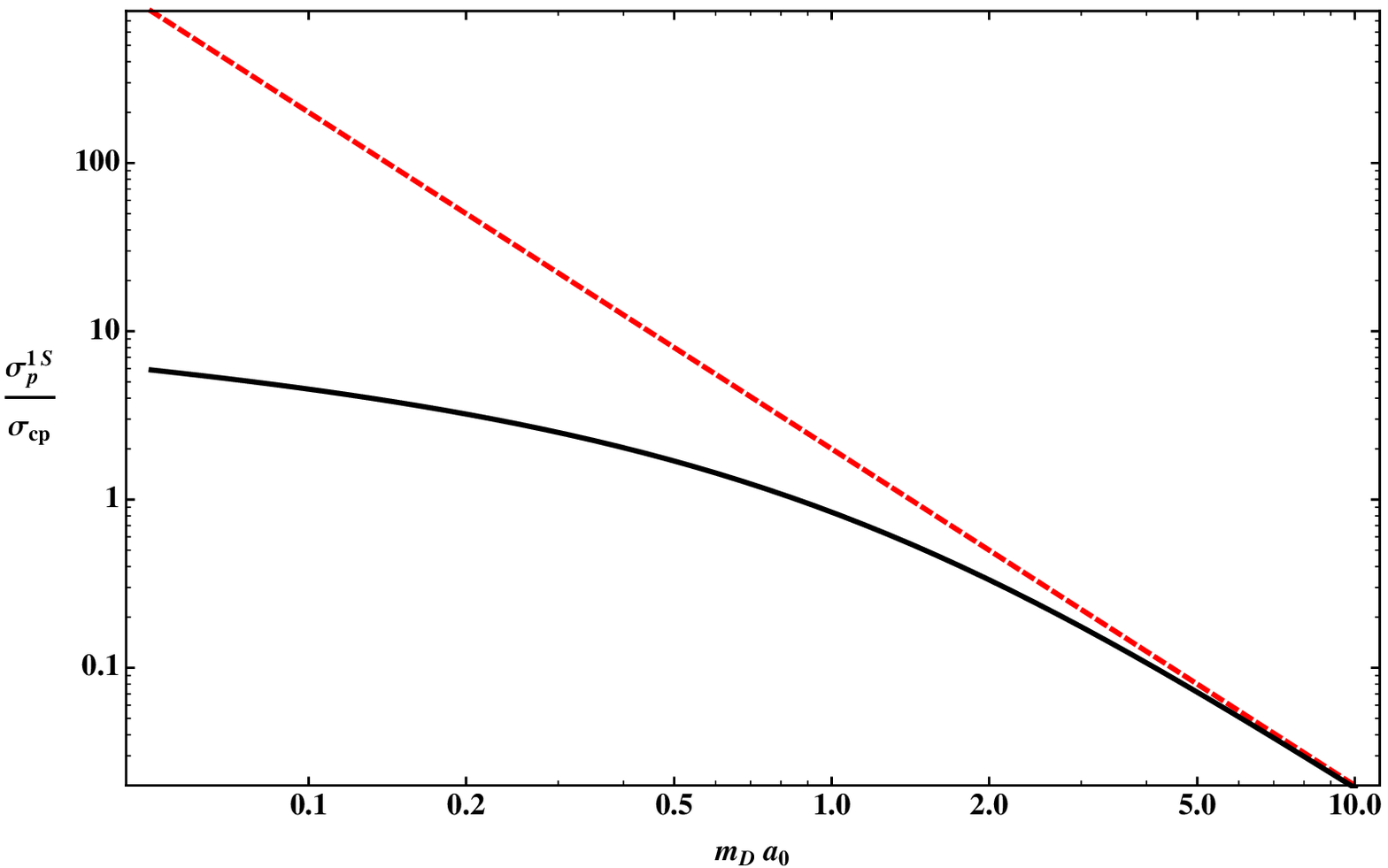}
\put(-160,145){\includegraphics[height=.035\textheight]{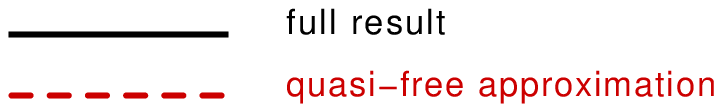}}
\caption{The cross section $\sigma^{1S}_p/\sigma_{cp}$ as a function of $m_Da_0$. The continuous line corresponds to
the exact result, whereas the red dashed line corresponds to the quasi-free approximation.}
\label{figqfa}
\end{figure}

The \emph{quasi-free approximation} amounts at replacing $\sigma_p^{nl}$ by $2\, \sigma_p^Q$, where  
$\sigma_p^Q$ is the in-vacuum cross section $p+Q \to p+Q$.
Hence the quasi-free approximation is equivalent to neglecting interference terms between 
the different heavy-quark lines in the amplitude square. Interference terms are the ones sensitive 
to the bound state. This corresponds in approximating $f(m_Da_0) \approx 2/(m_Da_0)^2$. 
The approximation holds only for $m_D \gg 1/a_0$, but at these temperatures quarkonium is dissociated.
For lower temperatures the approximation is largely violated by bound-state effects, see Figure~\ref{figqfa}.

\subsubsection{The temperature region $T\sim Mv\gg m_D$ }
In the temperature region $T\sim Mv\gg m_D$ the $1S$ dissociation cross section by parton scattering is 
\be
\sigma^{1S}_{p}=\sigma_{cp}\,h_p(m_Da_0,qa_0), 
\ee
where 
\bea
&& h_q(m_Da_0,qa_0) \equiv  -\ln\left(\frac{(m_Da_0)^2}{4}\right)-\frac{3}{2} +\ln\left(\frac{(qa_0)^2}{1+(qa_0)^2}\right) -\frac{1}{2(qa_0)^2}\ln(1+(qa_0)^2), 
\\
&& h_g(m_Da_0,qa_0) \equiv -\ln\left(\frac{(m_Da_0)^2}{4}\right)-\frac{3}{2}  +\ln\left(\frac{(qa_0)^2}{1+(qa_0)^2}\right) + \frac{1}{2(1+(qa_0)^2)} 
-\frac{1}{(qa_0)^2}\ln(1+(qa_0)^2). 
\eea
The cross section is momentum dependent.

\begin{figure}[htb]
\includegraphics[height=.3\textheight]{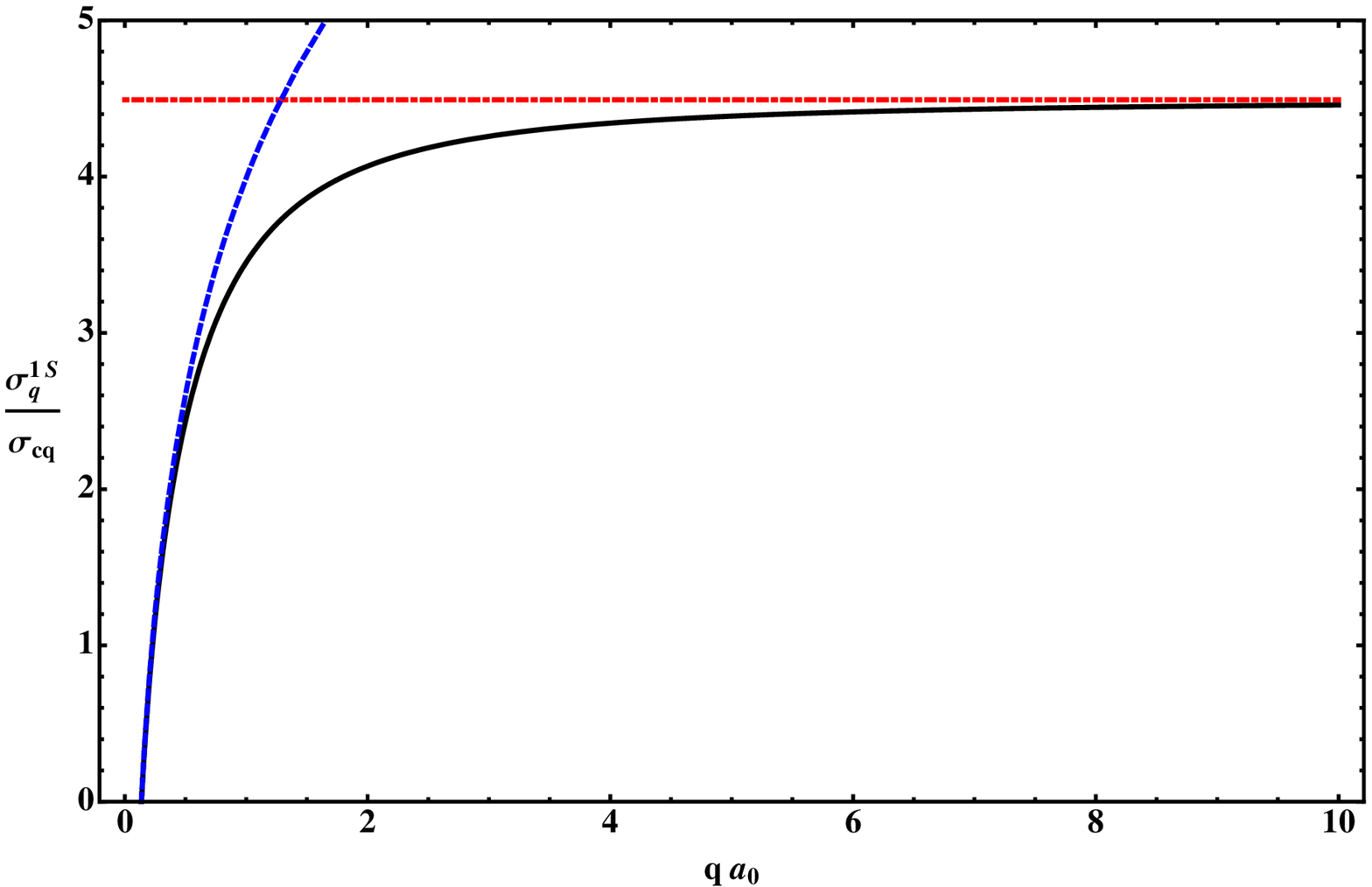}
\put(-180,50){\includegraphics[height=.1\textheight]{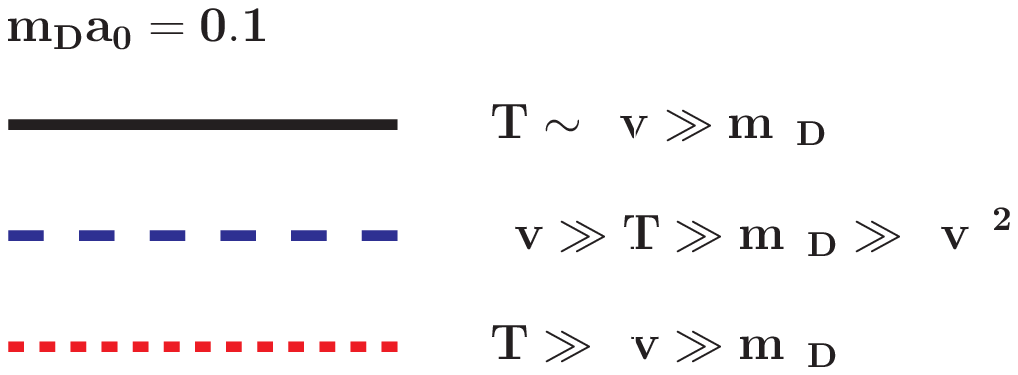}}
\caption{Dissociation cross sections due to scattering with light quarks, $\sigma^{1S}_q/\sigma_{cq}$, as a function of $q\,a_0$. 
  The dashed blue curve shows the cross section for $Mv\gg T\gg m_D\gg Mv^2$, 
  the continuous black curve shows the cross section for $T\sim Mv\gg m_D$, 
  and the dot-dashed red curve shows the cross section for $T\gg Mv \gg m_D$; for all the curves $m_D\,a_0=0.1$.}
\label{figquark}
\end{figure}

\begin{figure}[htb]
\includegraphics[height=.3\textheight]{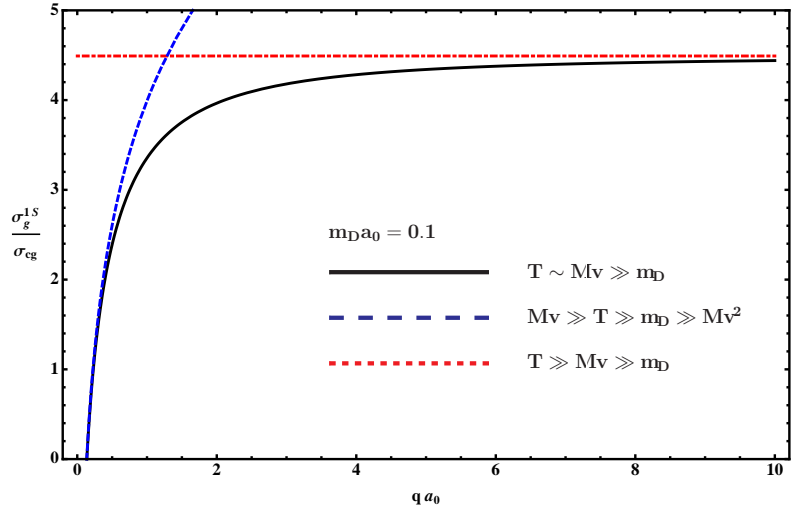}
\put(-180,50){\includegraphics[height=.1\textheight]{quasifreecomp_ins2_bis}}
\caption{Dissociation cross sections due to scattering with gluons, $\sigma^{1S}_g/\sigma_{cg}$, as a function of $q\,a_0$. 
  The dashed blue curve shows the cross section for $Mv\gg T\gg m_D\gg Mv^2$, 
  the continuous black curve shows the cross section for $T\sim Mv\gg m_D$,  
  and the dot-dashed red curve shows the cross section for $T\gg Mv \gg m_D$; for all the curves $m_D\,a_0=0.1$.}
\label{figgluon}
\end{figure}

\subsubsection{The temperature region $Mv\gg T \gg m_D \gg Mv^2$}
\begin{figure}[htb]
\makebox[-4.5truecm]{\phantom b}
\put(-30,0){\epsfxsize=10truecm \epsfbox{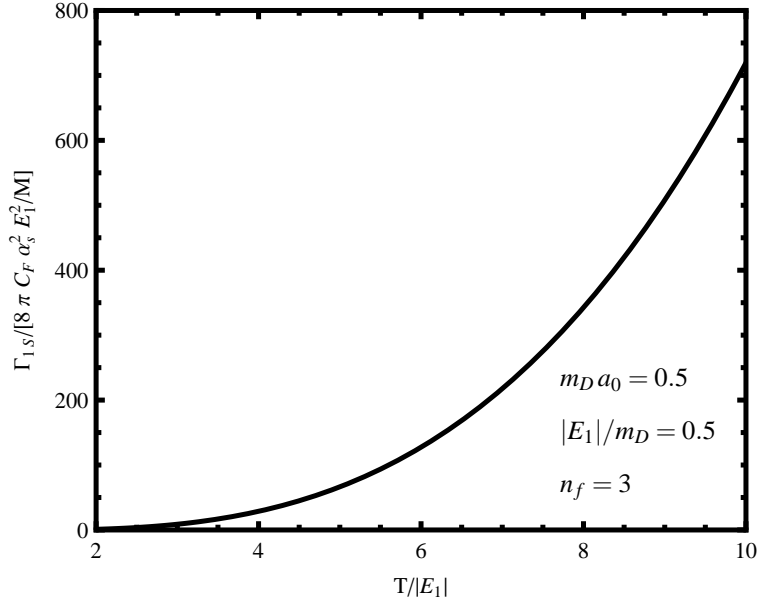}}
\put(180,80){$m_D\,a_0 = 0.5$}
\put(180,60){$|E_1|/m_D = 0.5$}
\put(180,40){$n_f = 3$}
\caption{Thermal width due to inelastic parton scattering. We have integrated over the continuous black curves shown 
in Figures~\ref{figquark} and \ref{figgluon} and summed over all partons according to \eqref{partonwidth}.}
\label{figwidth}
\end{figure}

In the temperature region  $Mv\gg T \gg m_D \gg Mv^2$ the $1S$ dissociation cross section by parton scattering is given by 
\be
\sigma^{1S}_p(q)=\sigma_{cp}\left[\ln\left(\frac{4q^2}{m^2_D}\right)-2\right].
\ee
The cross sections due to inelastic scatterings with quarks and gluons in the three different regimes 
are summarized in Figures~\ref{figquark} and \ref{figgluon} respectively~\cite{Brambilla:2013dpa}.
The thermal width follows from~\eqref{partonwidth} and is displayed in Figure~\ref{figwidth}~\cite{Vairo:2014xea}.

\subsubsection{The temperature region $Mv\gg T \gg Mv^2 \gg m_D$} 
In the temperature region $Mv\gg T \gg Mv^2 \gg m_D$, gluodissociation becomes such a competitive dissociation process  
to dominate, at least parametrically, over the dissociation by parton scattering.

In fact, the $1S$ dissociation cross section by parton scattering is in this regime 
\be
\sigma^{1S}_p(q)=\sigma_{cp}\left[\ln\left(\frac{4q^2}{m^2_D}\right)+\ln 2 -2\right],
\ee
which induces a thermal width of order $\als T \times (m_D/Mv)^2$.

The $1S$ gluodissociation cross section is at leading order 
\be 
\sigma^{1S}_{\rm gluo\, LO}(q)=\frac{\als\cf}{3} 2^{10} \pi^2  \rho  (\rho +2)^2 \frac{E_1^{4}}{Mq^5}
\left(t(q)^2+\rho ^2	\right)\frac{\exp\left(\frac{4 \rho}{t(q)}  
\arctan \left(t(q)\right)\right)}{ e^{\frac{2 \pi  \rho}{t(q)} }-1}, 
\label{sigmagluolo}
\ee
where $\rho\equiv 1/(\nc^2-1)$, $t(q)\equiv\sqrt{q/\vert E_1\vert-1}$ and $E_1 = - M\cf^2\als^2/4$~\cite{Brambilla:2011sg,Brezinski:2011ju}.
This cross section induces a thermal width of order $\als T \times (Mv^2/Mv)^2$, i.e., larger 
by a factor $(Mv^2/m_D)^2$ than the dissociation width by inelastic parton scattering. 
In this regime, gluodissociation is indeed the dominant dissociation mechanism.

\begin{figure}[htb]
\includegraphics[height=.35\textheight]{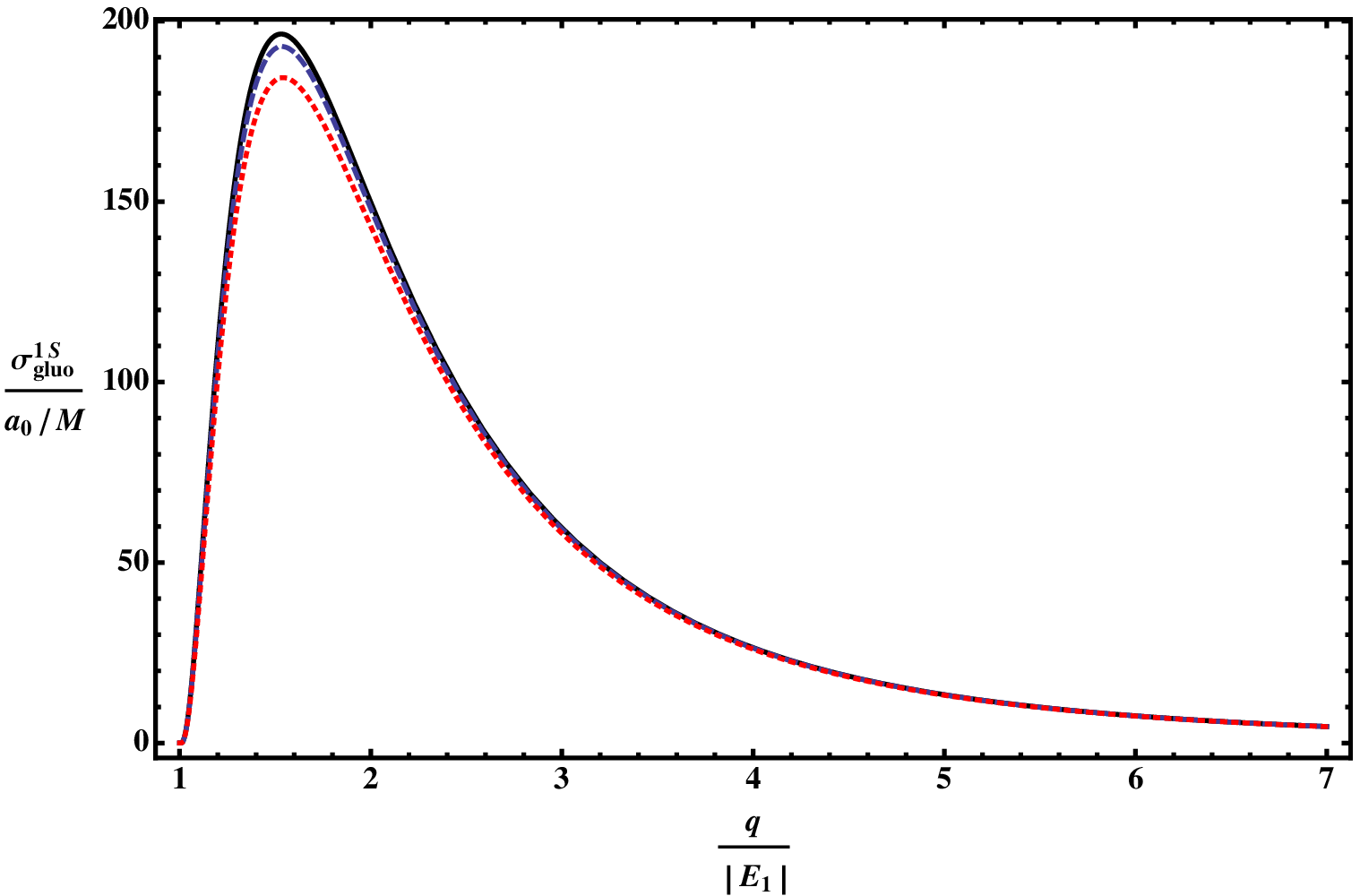}
\put(-170,150){\includegraphics[height=.07\textheight]{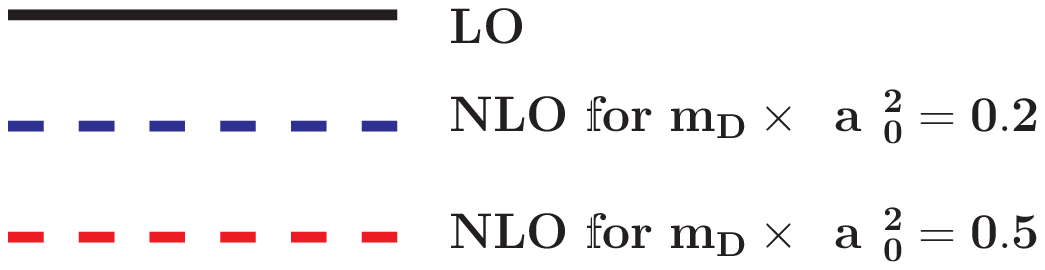}}
\caption{$1S$ gluo-dissociation cross section at leading order (continuous black line) 
and up to next-to-leading order for $m_D\,Ma_0^2=0.2$ (dashed blue line) and for $m_D\,Ma_0^2=0.5$ (dotted red line).}
\label{figlonlo}
\end{figure}

Of the same order as the dissociation by parton scattering is the gluodissociation at next-to-leading order, 
which has been calculated in~\cite{Brambilla:2013dpa}. 
The gluodissociation cross section up to next-to-leading order reads
\bea
&& \sigma^{1S}_{\rm gluo}(q)=  Z({q}/{m_D})\,\sigma_{\rm gluo\,LO}^{1S}(q),\\
&&  Z(q/m_D)=1-\frac{m_D^2}{4q^2}\left[\ln(8q^2/m_D^2)-2\right].
\eea
The result follows from adding a gluon self-energy diagram to the gluon propagator before or after the cut in Figure~\ref{figgluo}.
A comparison between the gluodissociation cross sections up to leading and next-to-leading order is in Figure~\ref{figlonlo}.

\begin{figure}[htb]
\makebox[0truecm]{\phantom b}
\includegraphics[height=.35\textheight]{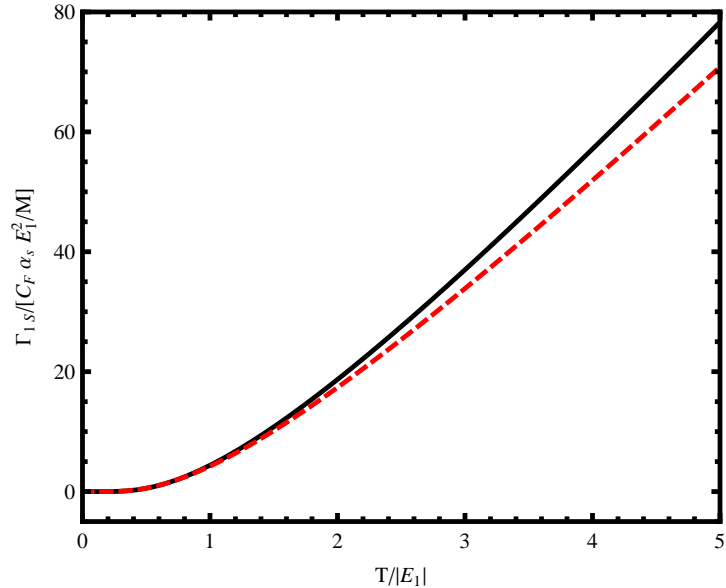}
\caption{Exact gluodissociation width at leading order (continuous black line) vs the Bhanot--Peskin approximation (dashed red line).}
\label{figcomp}
\end{figure}

The \emph{Bhanot--Peskin approximation} is a largely used approximation of the exact 
gluodissociation formula \eqref{sigmagluolo} that consists in neglecting final state interactions, 
i.e. the rescattering of a $Q\overline{Q}$ pair in a color octet configuration~\cite{Peskin:1979va,Bhanot:1979vb}.
Because the quark-antiquark colour-octet potential is $V_o = 1/(2\nc)\times \als/r$, it vanishes in the large $\nc$ limit:
\bea
\sigma^{1S}_{\rm gluo\, LO}(q) &\xrightarrow[\nc\to\infty]{}& 
16\frac{2^9\pi\als}{9}\frac{|E_1|^{5/2}}{m}\frac{(q+E_1)^{3/2}}{q^5} = 16 \,\sigma^{1S}_{\mathrm{BP}}(q),
\\
\Gamma_{1S\,{\rm LO}} &\xrightarrow[\nc\to\infty]{}& \int_{q \ge |E_1|}
\frac{d^3q}{(2\pi)^3} n_\mathrm{B}(q)\, 16 \,\sigma^{1S}_\mathrm{BP}(q) = \Gamma_{1S,\mathrm{BP}} \,,
\eea
where we have kept $C_F = 4/3$ in the overall normalization and $E_1 = - M\cf^2\als^2/4$.
The cross section $\sigma^{1S}_{\mathrm{BP}}$ and the width $\Gamma_{1S,\mathrm{BP}}$ are respectively 
the cross section and the width in the Bhanot--Peskin approximation. A comparison of the exact width 
with the  Bhanot--Peskin width is in Figure~\ref{figcomp}~\cite{Brambilla:2011sg}. 
At high temperatures, the exact width is about $13\%$ larger than its approximation, which is consistent with the large $\nc$ limit.
At low temperatures the thermal width goes to zero, however the exact and approximate widths 
vanish with different functional behaviours and the approximate width largely overshoots the exact one.

\section{Conclusions}
We have studied the dissociation of quarkonium in a thermal bath of gluons and light quarks 
in a framework that makes close contact with modern effective field theories for non relativistic bound states at zero temperature.
In a weakly-coupled framework, our conclusions may be concisely summarized in the following way.
\begin{itemize}
\item[$(i)$]{For $T < E_{\rm binding}$ the quark-antiquark potential in the medium coincides with the potential at $T=0$.}
\item[$(ii)$]{For $T > E_{\rm binding}$ the potential gets thermal contributions.}
\item[$(iii)$]{For $m_D < E_{\rm binding}$ quarkonium decays dominantly via gluodissociation (aka singlet-to-octet break up).} 
\item[$(iv)$]{For $m_D > E_{\rm binding} $ quarkonium decays dominantly via inelastic parton scattering (aka Landau damping).} 
\item[$(v)$]{For $T \simg T_{\rm dissociation} (< T_{\rm screening})$, quarkonium dissociates before forming.}
\end{itemize}
In a strongly-coupled framework, the hierarchy of non-relativistic scales is preserved, whereas 
the thermodynamical hierarchy may break down. This requires a non-perturbative definition and evaluation 
of the potential (real and imaginary).

\begin{theacknowledgments}
I acknowledge financial support from DFG and NSFC (CRC 110), 
and from the DFG cluster of excellence ``Origin and structure of the universe'' (www.universe-cluster.de).
\end{theacknowledgments}


\begin{thebibliography}{99}

\bibitem{Biondinihere} 
  S.~Biondini, talk at this conference.

\bibitem{Escobedohere} 
  M.~Escobedo, talk at this conference.

\bibitem{Brambilla:2008cx} 
  N.~Brambilla, J.~Ghiglieri, A.~Vairo and P.~Petreczky,
  Phys.\ Rev.\ D {\bf 78}, 014017 (2008)
  [arXiv:0804.0993 [hep-ph]].

\bibitem{Brambilla:2004wf}
  N.~Brambilla {\it et al.},
  CERN-2005-005, (CERN, Geneva, 2005)
  [arXiv:hep-ph/0412158].

\bibitem{Brambilla:2010cs} 
  N.~Brambilla, S.~Eidelman, B.~K.~Heltsley, R.~Vogt, G.~T.~Bodwin, E.~Eichten, A.~D.~Frawley and A.~B.~Meyer {\it et al.},
  Eur.\ Phys.\ J.\ C {\bf 71}, 1534 (2011)
  [arXiv:1010.5827 [hep-ph]].

\bibitem{Bazavov:2014pvz} 
  A.~Bazavov {\it et al.}  [HotQCD Collaboration],
  Phys.\ Rev.\ D {\bf 90}, 094503 (2014)
  [arXiv:1407.6387 [hep-lat]].

\bibitem{Matsui:1986dk} 
  T.~Matsui and H.~Satz,
  Phys.\ Lett.\ B {\bf 178}, 416 (1986).

\bibitem{Brambilla:2004jw} 
  N.~Brambilla, A.~Pineda, J.~Soto and A.~Vairo,
  Rev.\ Mod.\ Phys.\  {\bf 77}, 1423 (2005)
  [hep-ph/0410047].

\bibitem{Brambilla:2010vq} 
  N.~Brambilla, M.~A.~Escobedo, J.~Ghiglieri, J.~Soto and A.~Vairo,
  JHEP {\bf 1009}, 038 (2010)
  [arXiv:1007.4156 [hep-ph]].

\bibitem{Vairo:2010bm} 
  A.~Vairo,
  AIP Conf.\ Proc.\  {\bf 1317}, 241 (2011)
  [arXiv:1009.6137 [hep-ph]].

\bibitem{Caswell:1985ui} 
  W.~E.~Caswell and G.~P.~Lepage,
  Phys.\ Lett.\ B {\bf 167}, 437 (1986).

\bibitem{Bodwin:1994jh} 
  G.~T.~Bodwin, E.~Braaten and G.~P.~Lepage,
  Phys.\ Rev.\ D {\bf 51}, 1125 (1995)
  [Erratum-ibid.\ D {\bf 55}, 5853 (1997)]
  [hep-ph/9407339].

\bibitem{Pineda:1997bj} 
  A.~Pineda and J.~Soto,
  Nucl.\ Phys.\ Proc.\ Suppl.\  {\bf 64}, 428 (1998)
  [hep-ph/9707481].

\bibitem{Brambilla:1999xf} 
  N.~Brambilla, A.~Pineda, J.~Soto and A.~Vairo,
  Nucl.\ Phys.\ B {\bf 566}, 275 (2000)
  [hep-ph/9907240].

\bibitem{Kharzeev:1994pz} 
  D.~Kharzeev and H.~Satz,
  Phys.\ Lett.\ B {\bf 334}, 155 (1994)
  [hep-ph/9405414].

\bibitem{Xu:1995eb} 
  X.~M.~Xu, D.~Kharzeev, H.~Satz and X.~N.~Wang,
  Phys.\ Rev.\ C {\bf 53}, 3051 (1996)
  [hep-ph/9511331].

\bibitem{Grandchamp:2001pf} 
  L.~Grandchamp and R.~Rapp,
  Phys.\ Lett.\ B {\bf 523}, 60 (2001)
  [hep-ph/0103124].

\bibitem{Grandchamp:2002wp} 
  L.~Grandchamp and R.~Rapp,
  Nucl.\ Phys.\ A {\bf 709}, 415 (2002)
  [hep-ph/0205305].

\bibitem{Brambilla:2013dpa} 
  N.~Brambilla, M.~A.~Escobedo, J.~Ghiglieri and A.~Vairo,
  JHEP {\bf 1305}, 130 (2013)
  [arXiv:1303.6097 [hep-ph]].

\bibitem{Laine:2006ns} 
  M.~Laine, O.~Philipsen, P.~Romatschke and M.~Tassler,
  JHEP {\bf 0703}, 054 (2007)
  [hep-ph/0611300].

\bibitem{Escobedo:2010tu} 
  M.~A.~Escobedo and J.~Soto,
  Phys.\ Rev.\ A {\bf 82}, 042506 (2010)
  [arXiv:1008.0254 [hep-ph]].

\bibitem{Vairo:2014xea} 
  A.~Vairo,
  EPJ Web Conf.\  {\bf 71}, 00135 (2014)
  [arXiv:1401.3204 [hep-ph]].

\bibitem{Brambilla:2011sg} 
  N.~Brambilla, M.~A.~Escobedo, J.~Ghiglieri and A.~Vairo,
  JHEP {\bf 1112}, 116 (2011)
  [arXiv:1109.5826 [hep-ph]].

\bibitem{Brezinski:2011ju} 
  F.~Brezinski and G.~Wolschin,
  Phys.\ Lett.\ B {\bf 707}, 534 (2012)
  [arXiv:1109.0211 [hep-ph]].

\bibitem{Peskin:1979va} 
  M.~E.~Peskin,
  Nucl.\ Phys.\ B {\bf 156}, 365 (1979).

\bibitem{Bhanot:1979vb} 
  G.~Bhanot and M.~E.~Peskin,
  Nucl.\ Phys.\ B {\bf 156}, 391 (1979).

\end{thebibliography}
\end{document}